\documentclass[aps,superscriptaddress,twocolumn,showpacs,amsmath,amssymb]{revtex4}
\usepackage{graphicx}
\usepackage{dcolumn}
\usepackage{bm}

\newcommand{\g}{\hbar v_F}

\begin{document}

\title{Topological confinement in graphene bilayer quantum rings}

\author{L. Jader }\email{leandro_jader@fisica.ufc.br}
\author{J. M. Pereira Jr.}\email{pereira@fisica.ufc.br}\affiliation{Departamento de F\'isica, Universidade Federal do
Cear\'a, Caixa Postal 6030, Campus do Pici, 60455-900 Fortaleza,
Cear\'a, Brazil}
\author{Andrey Chaves}\email{andrey@fisica.ufc.br}\affiliation{Departamento de F\'isica, Universidade Federal do
Cear\'a, Caixa Postal 6030, Campus do Pici, 60455-900 Fortaleza,
Cear\'a, Brazil}\affiliation{Department of Physics, University of
Antwerp, Groenenborgerlaan 171, B-2020 Antwerp, Belgium}
\author{G. A. Farias}\email{gil@fisica.ufc.br}
\affiliation{Departamento de F\'isica, Universidade Federal do
Cear\'a, Caixa Postal 6030, Campus do Pici, 60455-900 Fortaleza,
Cear\'a, Brazil}
\author{F. M. Peeters}\email{francois.peeters@ua.ac.be}
\affiliation{Departamento de F\'isica, Universidade Federal do
Cear\'a, Caixa Postal 6030, Campus do Pici, 60455-900 Fortaleza,
Cear\'a, Brazil} \affiliation{Department of Physics, University of
Antwerp, Groenenborgerlaan 171, B-2020 Antwerp, Belgium}

\date{ \today }

\begin{abstract}

We demonstrate the existence of localized electron and hole states
in a ring-shaped potential kink in biased bilayer graphene. Within
the continuum description, we show that for sharp potential steps
the Dirac equation describing carrier states close to the $K$ (or
$K'$) point of the first Brillouin zone can be solved analytically
for a circular kink/anti-kink dot. The solutions exhibit
interfacial states which exhibit Aharonov-Bohm oscillations as
functions of the height of the potential step and/or the radius of
the ring.

\end{abstract}


\maketitle

Graphene, a one atom thick crystal sheet of carbon, has been shown
to display striking electronic and mechanical properties which are
expected to lead to the development of new devices (for a review,
see e.g. \cite{Castro1}). These properties are related to the
unusual electronic structure of graphene, in which the charge
carriers behave as massless fermions with a gapless linear
dispersion. For two coupled graphene sheets, known as bilayer
graphene (BG), the electronic structure is modified due to the
interlayer interaction, with the otherwise linear dispersion
becoming approximately parabolic. Another important feature of BG
is the fact that the electronic dispersion can develop a gap,
either by doping of one of the layers or by the application of an
external perpendicular electric field. Such a gap can be tuned by
varying the external electric field \cite{Falko}, which allows for
the possibility of tailoring the electronic structure of BG for
the development of devices, such as quantum dots
\cite{Milton1,Magdot}, and quantum rings \cite{Zarenia1,Zarenia2}.
Recently it has been shown that another consequence of the
existence of a tunable gap in BG is the possibility of topological
confinement of carriers in antisymmetric potential 'kinks', i.e.
at the interface between two regions of an antisymmetric external
electric field \cite{Morpurgo,Martinez,SanJose}. These states have
similarities with the surface states of topological insulators
\cite{topoinsul1,topoinsul2}. Their energies are found inside the
gap and the wavefunctions are predicted to decay away from the
interface of the kink potential. These topological states are
expected to be robust with respect to the effect of disorder, with
the carrier propagation along the potential kink displaying
electron-hole asymmetry. In addition, the geometry of the
potential interface is determined by the shape of the voltage
gates used to induce the gap. That in turn can be used to further
constrain the carrier propagation.

In this letter we propose a system in which topological confined
states are realized at the interface of a kink potential shaped as
a ring. This system can be regarded as a good approximation of an
ideal quantum ring of zero width. We obtain analytical expressions
for the electronic wavefunction of the BG four-band Hamiltonian.
Bilayer graphene can be described as two bipartite coupled sheets
with four triangular sublattices labeled as $A$ ($A'$) and $B$
($B'$) in the upper (lower) layer. For Bernal stacking
\cite{Milton2} the coupling between layers is described by the
hopping energy $\overline{t}$ = 400 meV between sites $A$ and
$B'$, so that the Hamiltonian around the $K$ point of the first
Brillouin zone can be written as
\begin{equation}
H = \left[\begin{array}{cccc}
\overline{U}_1 & \pi & \overline{t} & 0 \\
\pi^\dag & \overline{U}_1 & 0 & 0 \\
\overline{t} & 0 & \overline{U}_2 & \pi^\dag \\
0 & 0 & \pi & \overline{U}_2
\end{array}\right], \label{hamiltonian}
\end{equation}
where $\overline{U}_1$ and $\overline{U}_2$ are external
electrostatic potentials applied respectively to the upper and
lower layers. In polar coordinates we have $\pi= -i\g
e^{i\theta}\left(\frac{\partial}{\partial{\rho}}
+\frac{i}{\rho}\frac{\partial}{\partial{\theta}}\right)$ and
$\pi^\dag= -i\g e^{-i\theta}\left(\frac{\partial}{\partial{\rho}}
-\frac{i}{\rho}\frac{\partial}{\partial{\theta}}\right)$. The
eigenstates of the Hamiltonian Eq. (\ref{hamiltonian}) are four
component pseudo-spinors $\Psi = \left[ \psi_A, \quad \psi_B,
\quad \psi_{B'}, \quad \psi_{A'} \right]^T$, where $\psi_{A}$ and
$\psi_{B}$ ($\psi_{A'}$ and $\psi_{B'}$) are the envelope
functions for the probability amplitudes for sublattices $A$
($A'$) and $B$ ($B'$), respectively, in the upper (lower) layer.
The resulting four coupled differential equations can be
decoupled, giving, for $\overline{U}_2 = - \overline{U}_1$
\begin{eqnarray}
&&\nabla^2\nabla^2\psi_A+2(\epsilon^2+U_1^2)\nabla^2\psi_A+
\cr&&\cr &&+[(U_1^2-\epsilon^2)^2+t^2(U_1^2-\epsilon^2)]\psi_A=0,
\label{eqdecoup}
\end{eqnarray}
with $U_1 = \overline{U}_1/\g$, $U_2 = \overline{U}_2/\g$, $t =
\overline{t}/\g$ and $\epsilon = E/\g$, where $E$ is the energy.

\begin{figure}
\centering
\includegraphics[height = 7.0 cm]{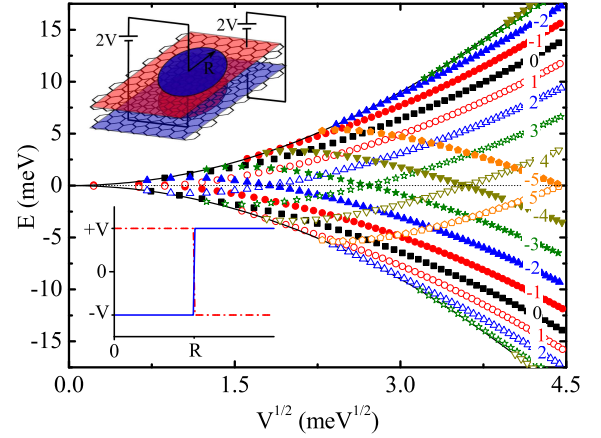}
\caption{ (Color online) Energy states with angular momentum index
$m = 0, \pm 1, ..., \pm 5$ as a function of the square root of the
kink/anti-kink potential height (lower inset) for a bilayer
graphene ring as schematically shown in the upper inset,
considering a ring radius $R = 50$ nm. Full (empty) symbols refer
to negative (positive) $m$. The black solid lines depicts the
limits $\pm V$ between the confined states and the continuum,
whereas the thin dotted line is $E = 0$.} \label{fig1}
\end{figure}

For the case of a quantum ring with radius $R$, following Ref.
\cite{Morpurgo}, we assume a sharp potential kink, so that one can
define two regions, namely \textbf{I}) $0\le \rho < R$ and
\textbf{II}) $\rho > R$. The upper inset of Fig. \ref{fig1}
illustrates a sketch of the system considered in the present
letter. In order to simplify the calculations, the potentials
$U_{1}$ (red, dashed-dotted) and $U_2$ (blue, solid) are assumed
to be piecewise constant, defined as $V$ and $-V$ for region
\textbf{I} and $-V$ and $V$ for region \textbf{II}, respectively,
as illustrated in the lower inset of Fig. 1. Notice that this
quantum ring potential is different from the one studied in Ref.
\cite{Zarenia1}, where the potential was defined as zero inside a
finite width circular ring and $U_1 = -U_2 = V$ otherwise. In
order to obtain an analytical solution for this problem, one must
solve the system of differential equations for each region and
match the eigenfunction at the boundaries. For region \textbf{I}
the solutions are Bessel functions, which obey the relation
$\nabla^2F_m(\alpha\rho)e^{im\theta}=\pm\alpha^2F_m(\alpha\rho)e^{im\theta}$,
where the function $F_m(\alpha\rho)$ denotes the Bessel functions
$J_m(\alpha\rho)$ or $Y_m(\alpha\rho)$, with eigenvalue
$+\alpha^2$, or the modified Bessel functions $I_m(\alpha\rho)$ or
$K_m(\alpha\rho)$, with eigenvalue $-\alpha^2$. The circular
symmetry of the problem implies that $m = 0, \pm 1, \pm 2, ...$.
By substituting the solutions one finds a fourth order algebraic
equation for $\alpha$, whose solutions are
\begin{equation}
\alpha=\pm\sqrt{V^2+E^2\pm\sqrt{4V^2E^2-t^2(V^2-E^2)}}
\end{equation}

Henceforth, we consider only energies within the interval
where $\alpha$ is complex, such that
$F_m(\alpha\rho)$ exhibits real and imaginary parts.
If one chooses
$\alpha=\sqrt{V^2+E^2-\sqrt{4V^2E^2-t^2(V^2-E^2)}}$,
four linearly independent (LI) solutions are found:
$\Re[J_m(\alpha\rho)]e^{im\theta}$,
$\Im[J_m(\alpha\rho)]e^{im\theta}$,
$\Re[K_m(i\alpha\rho)]e^{im\theta}$ and
$\Im[K_m(i\alpha\rho)]e^{im\theta}$. 
It can be verified
that these functions are solutions of Eq. (\ref{eqdecoup}),
although they are not separately eigenfunctions of the Laplacian.
It is important to point out that, for this choice of $\alpha$,
this set of functions is LI only when the imaginary part of
$\alpha$ is negative.

The functions $\Re[J_m(\alpha\rho)]e^{im\theta}$ and
$\Im[J_m(\alpha\rho)]e^{im\theta}$ are finite at the origin, but
diverge as $\rho\rightarrow\infty$ for any value of $m$. On
the other hand, $\Re[K_m(i\alpha\rho)]e^{im\theta}$ diverges for
$\rho\rightarrow0$ and vanishes when $\rho\rightarrow\infty$, for
any value of $m$. The function $\Im[K_m(i\alpha\rho)]e^{im\theta}$
has the same behavior as $\Re[K_m(i\alpha\rho)]e^{im\theta}$ for
$m\ne0$, but for $m=0$ it is finite at the origin. Based on this
we construct the wavefunction $\phi_A^<$ for the
region \textbf{I} as \cite{foot1}
\begin{equation}
\psi_{A}^<(\rho,\theta)=\{A\Re[J_m(\alpha\rho)]+B\Im[J_m(\alpha\rho)]\}e^{im\theta}.
\end{equation}

Using this expression for $\psi_A^<$, the radial part of the other
components of $\Psi^<$ for region \textbf{I} are
\begin{equation}
\phi_{B}^<(\rho)= \frac{i}{V-E}\{A\Re[\alpha
J_{m-1}(\alpha\rho)]+B\Im[\alpha J_{m-1}(\alpha\rho)]\},
\end{equation}
\begin{equation}
\phi_{B'}^<(\rho)= \frac{1}{t(V-E)}\{A\Re[\gamma_1
J_m(\alpha\rho)]+B\Im[\gamma_1 J_m(\alpha\rho)]\},
\end{equation}
\begin{eqnarray}
\phi_{A'}^<(\rho)&=& \frac{i}{t(V^2-E^2)}\{A\Re[\alpha
\gamma_1 J_{m+1}(\alpha\rho)]+\cr&&\cr &&B\Im[\alpha \gamma_1
J_{m+1}(\alpha\rho)]\},
\end{eqnarray}
where $\gamma_1 = \alpha^2-(V-E)^2$ and $\psi_{B}^<(\rho,\theta)=
\phi_{B}^<(\rho)e^{i(m-1)\theta}$, $\psi_{B'}^<(\rho,\theta)=
\phi_{B'}^<(\rho)e^{i(m)\theta}$, and $\psi_{A'}^<(\rho,\theta)=
\phi_{A'}^<(\rho)e^{i(m+1)\theta}$. Similarly, for region
\textbf{II}, we choose for the wavefunction $\psi_A^>$ a linear
combination of solutions that go to zero as $r \rightarrow
\infty$, namely,
\begin{equation}
\psi_{A}^>(\rho,\theta)=\{C\Re[K_m(i\alpha\rho)]+D\Im[K_m(i\alpha\rho)]\}e^{im\theta},
\end{equation}
and find the other components of $\Psi^>$ as
\begin{eqnarray}
\phi_{B}^>(\rho)&=& \frac{i}{(V+E)}\{-C\Im[
\alpha K_{m-1}(i\alpha\rho)]+\cr &&\cr &&D\Re[
\alpha K_{m-1}(i\alpha\rho)]\},
\end{eqnarray}
\begin{eqnarray}
\phi_{B'}^>(\rho)&=&
\frac{-1}{t(V+E)}\{C\Re[\gamma_2 K_m(i\alpha\rho)]+\cr &&\cr &&D\Im[
\gamma_2 K_m(i\alpha\rho)]\},
\end{eqnarray}
\begin{eqnarray}
\phi_{A'}^>(\rho)&=&
\frac{i}{t(V^2-E^2)}\{-C\Im[\alpha\gamma_2K_{m+1}(i\alpha\rho)]+
\cr &&\cr &&D\Re[\alpha\gamma_2K_{m+1}(i\alpha\rho)]\},
\end{eqnarray}
where $\gamma_2 = \alpha^2-(V+E)^2$ and $\psi_{B}^>(\rho,\theta) =
\phi_{B}^>(\rho)e^{i(m-1)\theta}$, $\psi_{B'}^>(\rho,\theta) =
\phi_{B'}^>(\rho)e^{i(m)\theta}$ and $\psi_{A'}^>(\rho,\theta) =
\phi_{A'}^>(\rho)e^{i(m+1)\theta}$. The continuity of the
wavefunction at $\rho = R$ implies that $\Psi^<(R, \theta) =
\Psi^>(R, \theta)$. That condition leads to a system of equations
from which we obtain the energy eigenvalues.

The energies of the topologically confined states of a bilayer
graphene ring with radius $R = 50$ nm are shown in Fig. \ref{fig1}
as function of the square root of the kink/anti-kink potential
height. The energy spectrum exhibits the symmetry $E(m) = -E(-m)$,
which corresponds to $E(k_y) = -E(-k_y)$ as found for a
one-dimensional kink/anti-kink potential in the $x$-direction
\cite{Morpurgo}. Notice that the $m = 0$ state is not the lowest
energy state, and the less energetic confined electrons in such a
system will have non-zero angular momentum index, even in the
absence of a magnetic field. The zero energy states are two-fold
degenerate (without taking spins into account), and are realized
only for specific values of the potential height $V$.

\begin{figure}
\centering
\includegraphics[height = 7.0 cm]{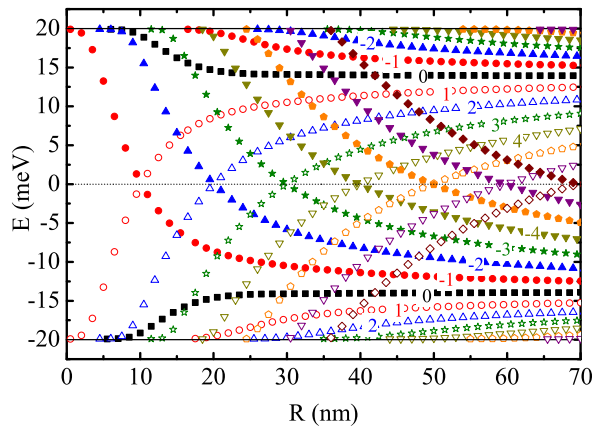}
\caption{ (Color online) Energy spectrum for $m = 0, \pm 1, ...,
\pm 7$ as a function of the ring radius for bilayer graphene in
the presence of a circular kink/anti-kink potential of height $V =
20$ meV. Full (empty) symbols refer to negative (positive) values
of $m$. Three lines are drawn to help visualization: the black
solid lines delimits the potential height $\pm V$, whereas the
thin dotted line is $E = 0$.} \label{fig2}
\end{figure}

Figure \ref{fig2} shows the energy states with angular momentum
index $m = 0, \pm 1, ..., \pm 7$  as a function of the ring radius
$R$ for a kink/anti-kink potential height $V = 20$ meV. The solid
lines delimits the -20 meV $< E <$ 20 meV energy spectrum of
confined states. Notice that double degenerate $E = 0$ states are
observed only for specific values of the radius. The energies
converge to two values, $\pm E_c$, as the ring radius increases,
forming two merged bands around these energies. The value $E_c
\simeq 13.9$ meV is identical to the energy of the corresponding
one-dimensional problem with $k_y = 0$.

\begin{figure}[!b]
\centering
\includegraphics[height = 7.0 cm]{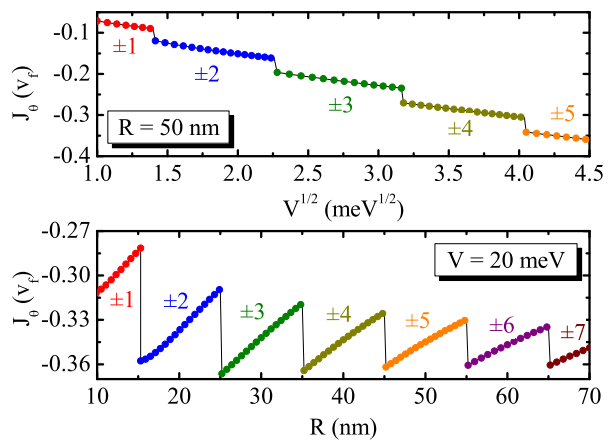}
\caption{ (Color online) Angular component of the probability
density current as a function of the square root of the
kink/anti-kink potential height $V$ (top), for a ring radius $R =
50$ nm, and as a function of $R$ (bottom), when $V = 20$ meV. The
number on each curve refers to the angular momentum index $m$, and
the solid curves are a guide to the eyes.} \label{fig3}
\end{figure}

The origin of the zero energy states in Figs. \ref{fig1} and
\ref{fig2} is similar to those found earlier for the
one-dimensional problem: the energy spectrum for a kink/anti-kink
potential, as shown in Ref. \cite{Morpurgo}, exhibits $E = 0$
states at two values of the linear momentum $\hbar k_y$ with the
same modulus, say, $\hbar k_y^{(0)}$ and $-\hbar k_y^{(0)}$, which
were shown to be proportional to the square root of the potential
height $V$ for $V \ll t$. In the bilayer graphene ring problem, an
analogy can be made between the angular momentum $L_z = \hbar m/R$
(see \emph{e. g.} Eq. (23) of Ref. \cite{Zarenia2}) and the linear
momentum $\hbar k_y$ of the one-dimensional case. Whenever $L_z =
\hbar m/R = \pm \hbar k_y^{(0)}$, zero energy states appear,
hence, if one fixes the potential height $V$, $k_y^{(0)}$ will be
a fixed value, and for each value of $m$, there will be a value of
$R$ that satisfies this condition, leading to a double degenerate
zero energy state for this value of the radius. For example, Fig.
\ref{fig2} shows the results for $V = 20$ meV, where $k_y^{(0)}$
is found to be $\sim 0.1$ nm$^{-1}$, consequently, zero energy
states are observed at $R \sim 10$ nm for $m = \pm 1$, $R \sim 20$
nm for $m = \pm 2$ and so on. Moreover, as the $E = 0$ states
satisfy the condition $m/R = k_y^{(0)} \simeq \sqrt{tV}/(2^{3/4})$
for $V \ll t$ \cite{Morpurgo}, the equally spaced zero energy
states observed in Fig. \ref{fig1} occur for $m/\sqrt{V} =
2^{-3/4}R\sqrt{t}$, which for the parameters of Fig. \ref{fig1}
becomes $\sqrt{V} \sim 0.9$ meV$^{1/2}$ for $m = \pm 1$, $\sqrt{V}
\sim 1.8$ meV$^{1/2}$ for $m = \pm 2$ and so on. It is worth to
point out that the condition $k_y^{(0)} = \sqrt{tV}/(2^{3/4})$
proposed by Martin \emph{et al.} \cite{Morpurgo} for the one
dimensional problem was obtained from the reduced 2 $\times$ 2
Hamiltonian, which is valid only for $V \ll t$, hence the
dependence of $k_y^{(0)}$ on $\sqrt{V}$ is no longer guaranteed
for large values of the kink/anti-kink potential height.

The fact that the energy of the lowest energy states oscillates as
function of $R$ and $V$, where angular momentum transitions take
place, suggests the possibility of observing persistent currents
in each valley $K$ or $K'$ induced by the external potential, in
analogy to Aharonov-Bohm rings, but in the absence of a magnetic
field. The angular component of the probability density current
\cite{Zarenia2} for the lowest energy state is shown as a function
of the kink/anti-kink potential $V$ (top) and the ring radius $R$
(bottom) in Fig. 3, where current jumps are observed when angular
momentum transitions occur between states with different $|m|$. We
point out that $J_{\theta}(m) = J_{\theta}(-m)$ and that the
current for $K$ and $K'$ valleys have opposite sign, so that the
net current, taking into account both valleys, is zero.

In summary, we demonstrated that confined quantum ring states can
be realized in a circular kink/anti-kink potential in bilayer
graphene. We obtained an analytical solution for the Dirac
equation describing electrons close to the Dirac point. Zero
energy states, with two-fold degeneracy, are realized for special
values of the radius and potential height. Angular currents for
the lowest energy state, which present oscillations due to angular
momentum transitions as $V$ or $R$ increases, are observed.
Although, for the sake of simplicity, the potential profile was
assumed to be abrupt in the present work, in a more realistic
description it should have a continuous shape. Nevertheless, the
present results must give at least a good qualitative agreement
with a kink/anti-kink BG ring, which would be helpful for the
understanding of future experiments on such a system.

This work was financially supported by CNPq, under contract
NanoBioEstruturas 555183/2005-0, FUNCAP, CAPES, the Bilateral
programme between Flanders and Brazil, the Belgian Science Policy
(IAP) and the Flemish Science Foundation (FWO-Vl).


\begin{references}
\bibitem{foot1} In the case $m=0$, we could naively include an
additional term $G\Im[K_0(i\alpha\rho)]$ in the expression for
$\phi_{A}^<$, as such a function does not diverge in the origin in
this case. However, if we calculate the function $\phi_{A'}^<$ for
the sublattice $A'$ in this case, we would find a term
$G\Im[K_1(i\alpha\rho)]$, which forces us to choose $G=0$, as this
function diverges in the origin.
\end{references}
\end{document}